\documentclass[a4paper,showpacs,prb,twocolumn]{revtex4}
\usepackage{amsfonts}
\usepackage{amsmath}
\usepackage{amssymb}
\usepackage{graphicx}

\begin{document}

\title{Light propagation in nanorod arrays}
\author{A. I. Rahachou and I. V. Zozoulenko}
\affiliation{Solid State Electronics, Department of Science and Technology, Link\"{o}ping
University 601 74, Norrk\"{o}ping, Sweden}
\date{\today }

\begin{abstract}
We study propagation of TM- and TE-polarized light in
two-dimensional arrays of silver nanorods of various diameters in
a gelatin background. We calculate the transmittance, reflectance
and absorption of arranged and disordered nanorod arrays and
compare the exact numerical results with the predictions of the
Maxwell-Garnett  effective-medium theory. We show that
interactions between nanorods, multipole contributions and
formations of photonic gaps affect strongly the transmittance
spectra that cannot be accounted for in terms of the conventional
effective-medium theory. We also demonstrate and explain the
degradation of the transmittance in arrays with randomly located
rods as well as weak influence of their fluctuating diameter. For
TM modes we outline the importance of skin-effect, which causes
the full reflection of the incoming light. We then illustrate the
possibility of using periodic arrays of nanorods as high-quality
polarizers.
\end{abstract}

\maketitle

\section{Introduction}

Resonance properties of nanoparticles have been observed for centuries
thanks to beautiful colors of gold- and silver-patterned stained glasses.
Over the last decade nanopatterned materials have attracted even increased
attention  due to their unique electronic and optical characteristics.
Novadays, they are considered as promising candidates for wide variety of
applications in subwavelength waveguiding \cite{2002_PRB_Mayer,
2004_EL_Quidant}, enhanced Raman scattering spectroscopy \cite%
{2005_PRB_Bachelier}, non-linear optics \cite{1989_APL_Haus}, photovoltaics
\cite{2000_SEMSC_Westphalen} and biological/medical sensing \cite%
{2004_NATBIO_Alivisatos} and many others.

A characteristic size of metallic nanoparticles $d$ is about an order of
magnitude smaller than the wavelength of incoming light $\lambda $, which
can excite collective oscillations of electron density inside the particle,
-  plasmons. The plasmon excitation results in an enhanced extinction
(extinction = absorption + scattering) as well as an increased intensity of
the electromagnetic field near the particle \cite{1995_book_Kreibig}.

The important issue that makes nanoparticles so attractive for sensing
applications is the effect of the geometry and size of nanoparticles and
surrounding environment on a position of the plasmonic resonance \cite%
{1995_book_Kreibig, 2005_OS_Khlebtsov, 2005_JPCB_Lee, 2003_JPCB_Kelly}. For
example, the presence of antibodies in cells affected by cancer modifies the
environment for the gold nanoparticles placed on a tissue and results in a
shift of extinction peak that can be easily imaged by conventional
microscopy \cite{2005_NL_Sayed}.

Recently it has also been demonstrated \cite{2004_JOSAB_Wang, 2003_JMO_Wang}
that embedding metallic nanoparticles into a polymeric matrix provides the
larger contrast in the effective refractive index of the blend material,
being much lower or higher than that of a pure polymer. Developing such the
materials can facilitate creating high-contrast-index photonic polymer
crystals.

Nanoparticles assembled in nanochains can also be applied as subwavelength
waveguides \cite{2004_EL_Quidant, 2003_NATMat_Maier}. In the case of closely
placed particles the coupling (and light propagation) arises from the
evanescent dipole field from each particle, which excites a plasmon on its
neighbour. This excitation travels along the chain, making the electron
density within all the particles oscillate in resonance.

In the present paper we will focus on light propagation in large arrays of
infinitely long nanorods. Prototypes of such the arrays have been recently
fabricated experimentally \cite{2006_NT_Dev, 2005_Nanotech_Losic}. These
arrays represent randomly oriented or aligned long rods (or spikes) of a
material (dielectric or metal), several tens of nanometers in diameter.
Despite of significant progress in nanofabrication technologies, to our
knowledge, however, the theoretical description of light propagation in
nanorod arrays is still missing.

The paper is organized as follows. In Section II we outline transmittance
properties of nanorod arrays within the framework of the Maxwell-Garnett
effective-medium theory. In Section III we present numerical modeling of
light propagation through periodic arrays of nanorods and compare the
results with the predictions of the Maxwell-Garnett theory. In Section IV
the effect of various types of disorder is studied.

\section{Effective medium theory}

\begin{figure}[!htb]
\begin{center}
\includegraphics[keepaspectratio,width=0.6\columnwidth]{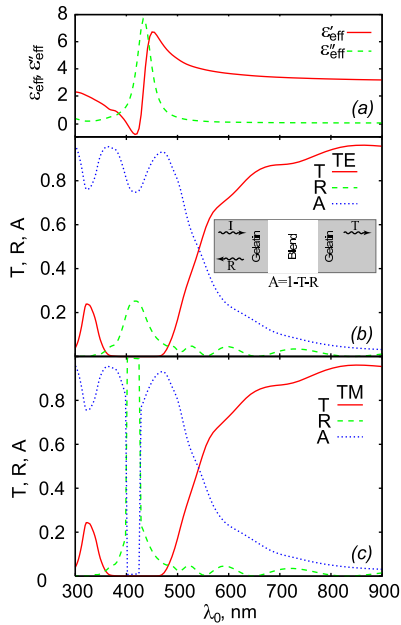}
\end{center}
\caption{(Color online) (a) Dielectric function of a blend of
silver nanorods (nanoparticles) with the concentration 10\%
embedded into a gelatin background. Transmittance, reflectance and
absorprtion of the TE (b) and TM (c) modes propagating through a 0.7$\protect%
\mu m$ thick layer of Ag(10\%)-gelatin blend. Inset in (b)
outlines the system under study.} \label{fig:eff_medium}
\end{figure}

We consider a gelatin matrix with an embedded two-dimensional array of
silver nanorods. The effective dielectric function $\varepsilon_{eff}(\omega)
$ of that composite can be estimated from developed for more than 100 years
ago Maxwell-Garnett theory \cite{1995_book_Kreibig}:

\begin{equation}
\frac{\varepsilon _{eff}(\omega )-\varepsilon _{mat}}{\varepsilon
_{eff}(\omega )+2\varepsilon _{mat}}=f\frac{\varepsilon _{rod}(\omega
)-\varepsilon _{mat}}{\varepsilon _{rod}(\omega )+2\varepsilon _{mat}},
\label{eq:maxwell_garnet}
\end{equation}%
where $f=S_{2}/S_{1}$ is the filling factor of the nanorods embedded into
the matrix, $S_{1}$ is the active area of the matrix and $S_{2}$ is the
total cross-section area of the nanorods. The dielectric function  of the
gelatin matrix is $\varepsilon _{mat}=2.25$. The dielectric function $%
\varepsilon _{rod}(\omega )$ of the nanorods is taken from the SOPRA
database \cite{www_sopra} for the bulk material. The Maxwell-Garnet theory
is valid for relatively small nanoparticles (nanorods) (up to several tens
of nanometers) at low concentrations (less then 30\%). The dielectric
function (here and hereafter all the spectra are given in respect of light
wavelength in vacuum $\lambda _{0}$) of the Ag(10\%)-gelatin blend is
presented in Fig. \ref{fig:eff_medium} (a).

The dielectric function in Fig. \ref{fig:eff_medium} (a) characterizes the
blend as a higly-dispersive lossy material with an absorption peak centered
around 414 nm. According to the Mie's theory this peak corresponds the
plasmon resonance of a single Ag spherical nanoparticle in gelatin, the
position of the peak obeys the well-known relation $\varepsilon_{rod}=-2
\varepsilon_{mat}$ \cite{1995_book_Kreibig}. In order to study light
propagation through the layer of the blend we consider a 2D
``sandwich-like'' structure consisting of semi-infinite gelatin
``waveguides'' connected to the blend region [see inset to Fig. \ref%
{fig:eff_medium} (b)]. The structure is assumed to be infinite in $z$%
-direction, thus the solution to the Maxwell's equations decouples into TE
(vector of a magnetic field is parallel to $z$) and TM (vector of an
electric field is parallel to $z$). The transmission, reflection and
absorption for both polarizations are given in Fig. \ref{fig:eff_medium} (b)
and (c) respectively.

It is easy to see that for both TE and TM polarizations there exists a gap
(or a stop-band) in the transmission caused by the enhanced absorption near
the extinction resonance peak. However, the reflectance and absorption
within the stop-band possess distinct behavior for different polarizations.
When the real part of the dielectric constant of the blend becomes negative (%
$400<\lambda _{0}<425$ nm) the reflectance of the TE mode increases due to
increased contrast against the dielectric function of the gelatin matrix
(which causes a dip in the absorption). At the same time, for TM-polarized
light the reflectance sharply increases up to 1 because of the metallic
character of the blend in this region and enhanced skin effect. For both
polarizations Bragg's reflections from the boundaries of the blend region,
manifested themselves as minima and maxima, are clearly seen for $\lambda
_{0}>500$ nm.

Despite its adequacy for small isolated circular nanoparticles, a simple
Maxwell-Garnet theory, however, has certain limitations. Namely, it does not
account for the shape and distribution of metal clusters in the dielectric
medium, neglecting important polarization properties of both single
non-circular particles and their arrangements \cite{2003_OC_Rechberger,
2006_ARXIV_Fung}. In order to incorporate these features and study
transmission characteristics of periodic and disordered nanorod arrays we
apply the recursive Green's function technique \cite{2005_PRB_Rahachou}.

\section{Periodic nanorod arrays}

\begin{figure}[!htb]
\begin{center}
\includegraphics[keepaspectratio,width=\columnwidth]{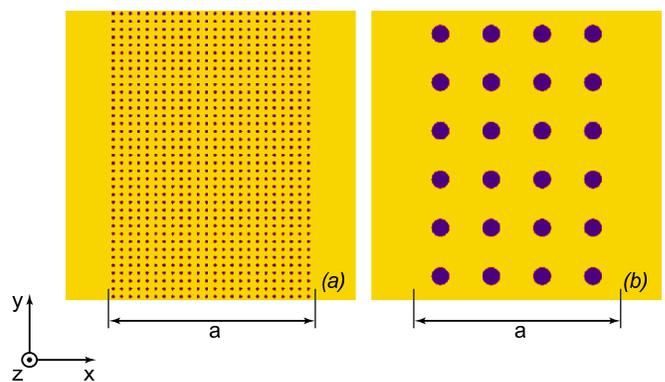}
\end{center}
\caption{(Color online) Arrays of silver nanorods with diameter
(a) 10 nm, and (b) 60 nm embedded in an infinite gelatin
background. For both cases the thickness of the layer $a=0.7
\protect\mu m$ and the filling factor $f$=10\%.}
\label{fig:nanorod_arrays}
\end{figure}

We now focus on 2D arrays of infinitely long silver nanorods arranged as a
square lattice in a gelatin background. Keeping the filling factor of Ag, $%
f=10\%,$ constant, we consider two cases, (a) a finite-size lattice with
thickness $a=0.7\mu m$ of nanorods with the diameter $d=10nm$, and (b) the
lattice of the same width assembled from nanorods of 60 nm in diameter, see
Fig. \ref{fig:nanorod_arrays}. Lattice constants are 29 and 175 nm for cases
(a) and (b) respectively.

Such the choice of nanorod sizes is motivated by the essential difference in
polarization properties of small and large nanoparticles \cite%
{2003_JPCB_Kelly}. If the nanoparticle is small enough ($d<<\lambda _{0}$),
according to the Mie's theory, only the dipole plasmonic oscillations
contribute to the extinction spectra, whereas for larger particles
higher-order resonances contribute to the spectra as well \cite%
{2003_JPCB_Kelly}. Using the recursive Green's function technique we perform
numerical simulations for both TE and TM polarizations of light falling
normally from the left to the boundary between gelatin and the blend.

\subsection{TE-modes}

When a nanosized metallic nanoparticle is illuminated by light, the electric
components of an electromagnetic field excite collective oscillations of
electronic plasma inside the particles -- plasmons. If the particles are
arranged into chains, these plasmonic oscillations possess a resonant
character that facilitates the propagation of light along the chain. Such
the chains have been intensively studied in literature \cite{2004_EL_Quidant}
as promising candidates for sub-wavelength wave-guiding.

Let us irradiate the array of infinitely long nanorods with TE-polarized
light. In this case $E_{x}$ and $E_{y}$ components of the electromagnetic
field excite coherent plasmonic oscillations on each nanorod. Figure \ref%
{fig:TE_modes_ordered} shows the calculated transmittance, reflectance and
absorption of a TE mode propagating through the arrays of nanorods.

\begin{figure}[!htb]
\begin{center}
\includegraphics[keepaspectratio,width=\columnwidth]{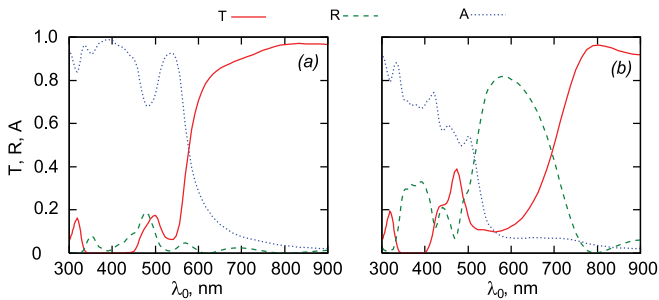}
\end{center}
\caption{(Color online) Transmittance, reflectance and absorption of a TE
mode travelling through the square arrays of nanorods with diameter (a) 10
nm and (b) 60 nm (see Fig. \protect\ref{fig:nanorod_arrays} for details).}
\label{fig:TE_modes_ordered}
\end{figure}

\emph{Small nanorods.} Let us first concentrate on an array of nanorods with
the diameter 10 nm [Fig. \ref{fig:TE_modes_ordered} (a)]. In the spectra one
can clearly distinguish two regions, namely the region of high absorption ($%
\lambda _{0}<600$ nm), containing a wide main absorption peak at 414 nm, two
minor peaks at 350 and 530 nm and the region of high transmittance ($\lambda
_{0}>600$ nm). Now we will take a closer look at these regions separately.

\begin{figure}[!htb]
\begin{center}
\includegraphics[keepaspectratio,width=\columnwidth]{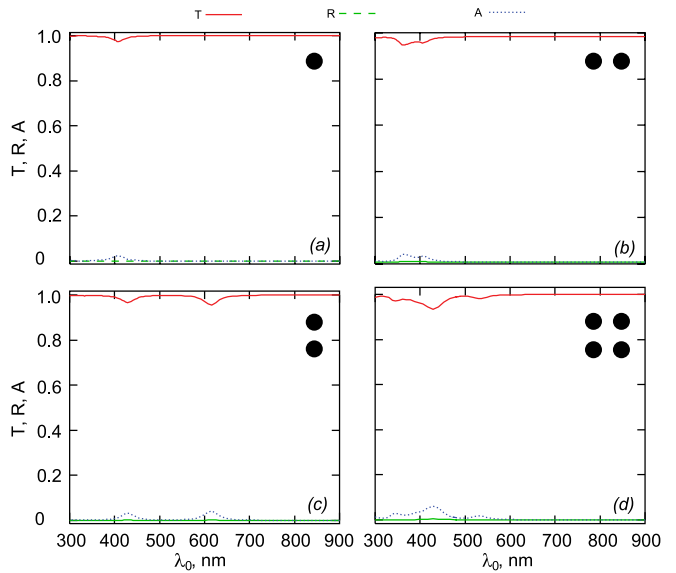}
\end{center}
\caption{(Color online) Transmittance, reflectance and absorption of a
single (a), a pair of horizontally (b) and vertically (c) aligned, and (d)
four coupled nanorods for the TE-polarized light. The inter-rod distances
are taken 29 nm, equal the lattice constant for the array (Fig. \protect\ref%
{fig:nanorod_arrays} (a)). The size of the computational domain is also the
same as that for the nanorod arrays in Fig. \protect\ref{fig:nanorod_arrays}.
}
\label{fig:single_nanoparticles}
\end{figure}

The position of the main extinction resonance agrees well with that obtained
from Eq. (\ref{eq:maxwell_garnet}). However, in contrast to the
Maxwell-Garnett theory, the spectrum contains two minor peaks near 350 and
530 nm. In order to explain them one needs to account the effect of coupling
between several nanorods. For this sake we compare light propagation through
(a) a single isolated nanorod (diameter 10 nm), (b) two coupled nanorods
aligned parallel to the light propagation direction, (c) those aligned
perpendicularly, and (d) four coupled nanorods. These four cases are
presented in Fig. \ref{fig:single_nanoparticles}.

For the case of a single isolated rod [Fig. \ref{fig:single_nanoparticles}
(a)] only one peak near 410 nm emerges which is in a good agreement with the
analytical value of 414 nm. For the cases (b) and (c) of twin coupled
nanorods the additional peaks, centered at 355 and 620 nm respectively,
appear. Their origin has been thoroughly studied in \cite{2003_OC_Rechberger}
and clarified in terms of enhanced [case (b)] and weaken [case (c)]
restoring forces between the particles. However, for the system of four
particles (d) these forces partially compensate each other, and the minor
resonances move closer towards the main peak.

\begin{figure}[!htb]
\begin{center}
\includegraphics[keepaspectratio,width=\columnwidth]{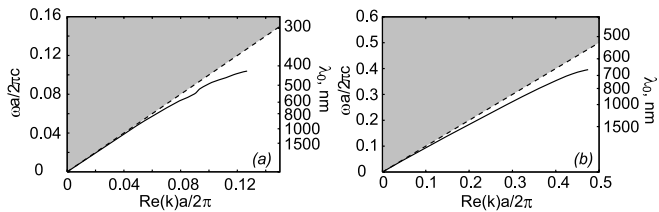}
\end{center}
\caption{Band diagrams of the nanorod arrays from Fig. \protect\ref%
{fig:nanorod_arrays} (a) and (b) respectively. The dashed line outlines the
light cone.}
\label{fig:disp_rel}
\end{figure}

Let us now focus on the wavelength region $\lambda _{0}>600$ nm, where
TE-polarized light propagates at high transmittance. In order to understand
this behavior, we complement the transmission coefficient with the band
diagram of the nanorod array. It should be mentioned that, in general, a
band diagram represents propagating Bloch states (states with real
eigenvalues). However, as the metallic rods (or nanoparticles) are
absorbing, all the states in the blend will be decaying and eventually die
off at the infinity. Yet all the Bloch eigenvalues in such systems have
imaginary components. In Fig. \ref{fig:disp_rel} (a) we represent a band
structure (real parts of eigenvalues) in $\Gamma X$-direction for the states
with the smallest imaginary parts.

The dispersion curve in Fig. \ref{fig:disp_rel} (a) has a small bump around
550 nm, which is caused by the minor extinction resonance. The band is
located very close to the light line that results in a rather strong
coupling between the incoming light and the plasmonic Bloch states of the
blend region. Such the strong coupling explains high transmittance in the
red wavelength region.

\emph{Large nanorods} For nanorods with the diameter 60 nm, the position of
the main extinction peak agrees with that one of the small particles.
However, there is an essential difference in physics behind. When the
diameter of a nanoparticle increases, higher-order dipole oscillations now
contribute the resulting extinction spectrum \cite{1995_book_Kreibig}. It
has been recently shown \cite{2003_JPCB_Kelly} that the peak centered at $%
\approx 400$ nm is due to the \emph{quadruple} resonance of a nanorod,
whereas the dipole resonance is redshifted and overlaps with the region of
the enhanced reflectance ($500<\lambda _{0}<700$ nm). The indication in
favor of this interpretation is a narrower width of the stop-band in the
transmission (60 nm against 100 nm in case of small rods). This is because
the higher-order dipole  interactions causing the stop-band behaviour for
the case of large nanorods are generally weaker.

Now let us clarify the origin of the high-reflectance region. The lattice
constant for this structure is 175 nm. This is of the same order as the
wavelength of light, such that the structure effectively represents a
two-dimensional photonic crystal. The plasmonic band in Fig. \ref%
{fig:disp_rel} (b) extends from $\omega a/2\pi c=0$ to 0.4 ($\lambda
_{0}\approx 660$ nm) where it experiences a \emph{photonic bandgap} that
causes the high reflectance of the structure. This bandgap overlaps with the
tail of the extinction peak near 500 nm (see Fig. \ref{fig:TE_modes_ordered}%
).

\subsection{TM-modes}

Let us now consider the TM-polarization of the incoming light. Figure \ref%
{fig:TM_modes_ordered} (a) shows the transmittance, reflectance and
absorption of the TM-polarized light for the small nanorods. In contrast to
the Maxwell-Garnett picture (Fig. \ref{fig:eff_medium}), almost for the
whole wavelength range under study light does not penetrate the region
occupied by nanorods and gets fully reflected back, resulting in zero
transmittance. This discrepancy can be explained by the skin-effect on
the silver rods. At the same time, the Maxwell-Garnett theory (\ref%
{eq:maxwell_garnet}) fully disregards the important screening properties of
the rods, simply averaging the effective dielectric constant over the
structure. It is also worth mentioning, that as we consider infinitely long
nanorods, incoming TM-mode \emph{does not excite} any plasmons on the rods
and thus there is no plasmonic contribution in overall transmission.

\begin{figure}[tbh]
\begin{center}
\includegraphics[keepaspectratio,width=0.6\columnwidth]{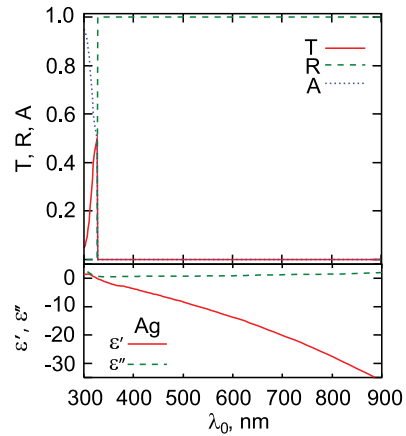}
\end{center}
\caption{(Color online) (a) Transmission, reflection and absorption
coefficients of the TM-mode through a nanorod array of $d=10$ nm. Due to the
skin-effect light does not penetrate the blend region. For $\protect\lambda %
_{0}<328$ nm the real part $\protect\varepsilon ^{\prime }$ of the
dielectric function of silver (b) becomes positive and the transmission
coefficient abruptly increases.}
\label{fig:TM_modes_ordered}
\end{figure}
However, for very short wavelengths ($\lambda _{0}\lesssim 328$ nm) the real
part of the dielectric function of silver becomes positive (see Fig. \ref%
{fig:TM_modes_ordered} (b)) and the blend behaves like a lossy dielectric
rather than a metal. This results in non-zer transmission in this region.

The obtained results clearly show that resonant plasmonic oscillations in
periodic nanorod arrays represent a dominating light propagation mechanism
for the TE-polarized light, whereas for the TM modes the nanorod structure
represents practically a perfect screen. This features can be utilized in a
nearly 100\% effective polarizer.

\section{Disordered nanorod arrays}

\begin{figure}[!htb]
\begin{center}
\includegraphics[keepaspectratio,width=\columnwidth]{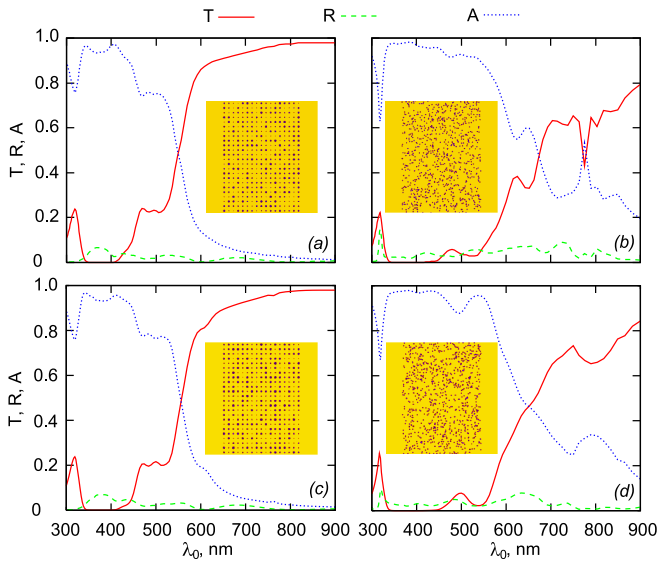}
\end{center}
\caption{(Color online) Transmission, reflection and absorption coefficients
of the TE-polarized light propagating through disordered nanorod arrays. (a,
c) Two different configurations of nanorods arranged in a square lattice,
their $d$ diameter randomly varies from of 5 to 20 nm. (b, d) Nanorods with
fixed $d=10$ nm are randomly distributed within the layer. Insets show the
actual geometries of the structures.}
\label{fig:disordered_str}
\end{figure}

As we have demonstrated in the preceding section, TE-polarized incident
light in off-resonance wavelength region propagates through periodic arrays
of small nanorods at very high transmission. Now we introduce some disorder
in this array and consider two separate cases, namely when nanorods are
arranged in a square lattice but have randomly varying diameter, and rods of
equal diameter, randomly distributed within the layer. For both cases the
filling factor $f=$10\% is kept constant, the distribution is taken uniform. Figures \ref{fig:disordered_str}
(a, c) and (b, d) demonstrate the transmittance, reflectance and absorbtion
for both cases.

The transmission characteristics for the structure with the random diameter
of nanorods [Fig. \ref{fig:disordered_str} (a, c)] closely resemble those
for the array of fixed-sized nanoparticles [Fig. \ref{fig:TE_modes_ordered}
(a)]. The main difference is that weaker dipole interactions between
adjacent particles of different diameter cause a minor narrowing of the
stop-band (60 nm versus 100 nm in the ideal case) and a slight degradation
of the minor extinction peak. It should be emphasized that the transmission
properties of arrays with different distributions of the nanorod diameter
[Fig. \ref{fig:disordered_str} (a) and (c)] are virtually the same.

For the case of the randomly distributed equal-sized nanorods [Fig. \ref%
{fig:disordered_str} (b, d)] the situation changes. In contrast to the
previous case of the ordered nanorod array with random diameter, the
absorption spectra in the region $\lambda _{0}>600$ nm are extremely
sensitive to the geometry of the structure. It is interesting to note how
clustering of nanorods manifests itself. The overall absorption in the
region $\lambda _{0}>600$ nm is much higher (and the transmision is lower)
in comparison to the periodic lattice, as it consists of the averaged
multiple absorbtion peaks of closely situated, touching or overlapping
nanorods. Since the inter-rod distances are not constant any longer, each
single rod is now affected by many dipole interactions of different
strengths from neighboring rods. Reflectances, however, are practically
identical and not significantly higher than for the periodic case. It can be
explained that due to its non-periodicity this structure absorbs better than
reflects. The clustering and more complex interactions of nanorods influence
the region $\lambda _{0}<600$ as well. The main and minor absorption peaks
for the structure Fig. \ref{fig:disordered_str} (b) almost overlap, whereas
for Fig. \ref{fig:disordered_str} (d) are still well-separated.

We should specially mention that in order to incorporate the effect of
nanoparticle aggregates into Maxwell-Garnett theory, several approaches were
suggested \cite{1995_book_Kreibig} (see also references therein). In that
case the effective medium dielectric function is derived by inserting the
total aggregate polarizabilities instead of that of a single isolated
particle into the Maxwell-Garnett theory.

\section{Conclusions}

We have studied propagation of TE- and TM-polarized light in
two-dimensional arrays of silver nanorods in a gelatin background.
In order to calculate transmittance, reflectance and absorption in
arrays of ordered and disordered nanorods we applied the recursive
Green's function technique and compared the obtained numerical
results with predictions of the Maxwell-Garnett effective-medium
theory. We have demonstrated that this theory describes adequately
only the case of the TE-polarized light propagating in ordered
arrays of small ($\sim 10$ nm), well-separated nanorods and only
in the frequency interval outside the main plasmonic resonance.

Our numerical calculations outline the importance of geometrical
factors such as the size of the rods and their distribution. In
particular, we have demonstrated, that interaction between
adjacent nanorods brings the significant contribution to the
transmission spectra, which is manifested as additional absorption
peaks (that are missing in the effective-medium approach). The
Maxwell-Garnett theory also disregards both the impacts of
higher-order dipole contributions and formation of photonic band
gaps in the case of arrays of large nanorods.

We have also studied the effect of disorder on the transmittance
of the nanorod arrays. We have introduced two types of disorder,
(a) ordered array with randomly varying nanorod diameters, and (b)
a random distribution of nanorods of the same size within the
blend. The disorder in rod placement leads to a strong suppression
of the transmission (and the enhanced absorption) due to plasmonic
resonances related to the clustering of the rods. We have
demonstrated that clustering effects are sensitive to the actual
geometry of the structure. In contrast, the impact of randomly
varying diameters of the rods is much less profound.

Despite its partial adequacy for the TE-polarized light, the
Maxwell-Garnett effective-medium theory is shown to be invalid for
the case of TM polarization. It simply averages the effective
dielectric function inside the blend, missing the important
screening properties of the metallic nanorods and characterizing
the blend as a (partially) transparent medium. In contrast, the
numerical modelling shows the strong skin effect that fully
prohibits the propagation of the TM modes through the structure.
The region of high transmittance for the TE modes and the strong
skin effect for the TM modes makes the nanorod arrays promising
candidates for high-quality polarizers.

\begin{acknowledgments}
We would like to thank Olle Ingan\"{a}s for stimulating and fruitful
discussions. We acknowledge a partial financial support from the Center for
Organic Electronics at Link\"{o}ping university. Useful conversations with
Nils-Christer Persson are also appreciated.
\end{acknowledgments}

\end{document}